\documentclass[letterpaper, preprint]{aastex}
\usepackage{amsmath, amsfonts, bm, calc}
\usepackage{graphicx}
\newcounter{address}
\newcommand{\latin}[1]{\emph{#1}}
\newcommand{\etal}{\latin{et\,al.}}

\newcommand{\eg}{\latin{e.\,g.}}
\newcommand{\unit}[1]{\mathrm{#1}}

\begin{document}
\title{
  An Affine-Invariant Sampler\\
  for Exoplanet Fitting and Discovery in Radial Velocity Data
}
\author{
  Fengji~Hou\altaffilmark{\ref{CCPP},\ref{email}},
  Jonathan~Goodman\altaffilmark{\ref{Courant}},
  David~W.~Hogg\altaffilmark{\ref{CCPP},\ref{MPIA}},
  Jonathan~Weare\altaffilmark{\ref{Courant},\ref{Chicago}},
  Christian~Schwab\altaffilmark{\ref{Yale},\ref{LSW}}
}

\setcounter{address}{1}
\altaffiltext{\theaddress}{\stepcounter{address}\label{CCPP} Center
  for Cosmology and Particle Physics, Department of Physics, New York
  University, 4 Washington Place, New York, NY 10003}
\altaffiltext{\theaddress}{\stepcounter{address}\label{email} To whom
  correspondence should be addressed: \texttt{fh417@nyu.edu}}
\altaffiltext{\theaddress}{\stepcounter{address}\label{Courant}
  Courant Institute of Mathematical Sciences, New York University, 251
  Mercer Street, New York, NY 10012}
\altaffiltext{\theaddress}{\stepcounter{address}\label{MPIA}
  Max-Planck-Institut f\"ur Astronomie, K\"onigstuhl 17, D-69117
  Heidelberg, Germany}
\altaffiltext{\theaddress}{\stepcounter{address}\label{Chicago}
  Department of Mathematics, University of Chicago, 5734 S.~University Avenue,
  Chicago, IL 60637}
\altaffiltext{\theaddress}{\stepcounter{address}\label{Yale}
  Department of Astronomy, Yale University, 260 Whitney Ave,
  New Haven, CT 06511}
\altaffiltext{\theaddress}{\stepcounter{address}\label{LSW}
  Universit\"at Heidelberg, Landessternwarte K\"onigstuhl 12, D-69117
  Heidelberg, Germany}

\begin{abstract}
Markov Chain Monte Carlo (MCMC) proves to be powerful for Bayesian inference
and in particular for exoplanet radial velocity fitting because MCMC
provides more statistical information and makes better use of data than
common approaches like chi-square fitting.  However, the non-linear density
functions encountered in these problems can make MCMC time-consuming.  In this
paper, we apply an ensemble sampler respecting affine invariance to orbital
parameter extraction from radial velocity data.  This new sampler has only
one free parameter, and it does not require much tuning for good performance,
which is important for automatization.  The autocorrelation time of this
sampler is approximately the same for all parameters and far smaller than
Metropolis-Hastings, which means it requires many fewer function calls to
produce the same number of independent samples.  The affine-invariant sampler
speeds up MCMC by hundreds of times compared with Metropolis-Hastings in the
same computing situation.  This novel sampler would be ideal for projects
involving large datasets such as statistical investigations of planet
distribution.  The biggest obstacle to ensemble samplers is the existence of
multiple local optima; we present a clustering technique to deal with local
optima by clustering based on the likelihood of the walkers in the ensemble.
We demonstrate the effectiveness of the sampler on real radial velocity data.

\end{abstract}


\section{Introduction}

Markov Chain Monte Carlo has proven very helpful in astrophysics for searching,
optimizing, and sampling probability distributions \citep{brooks11a}.  In most
cases, astrophysicists use the Metropolis--Hastings \citep{gilks96} algorithm
with a carefully tuned proposal distribution.  Those who have departed from
Metropolis--Hastings often have done so not to improve speed but to improve
the properties of the sampler for computing the Bayesian evidence integral
\citep{gregory92a, loredo99a, wandelt04a}.

In particular, MCMC has proven to be powerful for identification of exoplanet
signals in radial velocity (RV) data, and to quantify the reliability and
accuracy of exoplanet parameter inferences \citep[\eg,][]{ford05a, driscoll06a,
ford06b}.  MCMC produces samples of the posterior distribution of the orbital
parameters of each exoplanet around its target star \citep{gregory05a,
ford06a, ford06b}.  In turn, the sampling permits trivial (approximate)
posterior probability calculations and parameter marginalization, to obtain
the distribution of individual parameter or the joint distribution of
several parameters.  The standard Metropolis or Gibbs sampling approaches
have several practical difficulties.  One is that they have problem-specific
computational parameters, and possibly a large number of them, especially
those highly correlated, that require hand tuning \citep{ford06a, gregory11a}.
Even with optimally tuned diagonal parameters, these methods can be very slow
on some datasets. And of course, generically, there are local optima in the
likelihood function that correspond to incorrect companion identifications,
see \figurename~\ref{fig:loglikelihoodcdf}.

In this paper, we apply a new MCMC method \citep{gilks94, goodman10a}---an
ensemble sampler---to radial-velocity analysis.  This ensemble sampler has
the property of being invariant under affine transformations (see below for
precise definitions).  This means that the new sampler automatically, and
without hand tuning, works in the optimal linear rescaling of the problem.
This results in much smaller autocorrelation times, which means that one gets
the same amount of information about the posterior with fewer evaluations
of the likelihood function.  Our code has two other features that proved
indispensable in analyzing exoplanet RV data.  The first is a simulated
annealing cooling schedule that we use to generate the initial sample.
The second feature is a simple clustering method that identifies and removes
samples from irrelevant local optima.

\section{Model}

Our goal is to determine what configurations of planets around a
target star are consistent with given radial velocity
measurements.  The physical model consists of a collection of
Keplerian orbits.  The orbit of each planet is parameterized by five
orbital parameters \citep{ford06a, ford06b, gregory05a, ohta05a}.
This is consistent with the approximation that on the observational time
scale of a few years, interactions between the orbits are negligible
otherwise the orbits would not be stable.
This leads to
\begin{equation}
v_{\mbox{\scriptsize \em rad}}(t)=v_0+\sum_p \Delta v_p(t) \; ,
\end{equation}
where $v_0$ is a constant velocity offset, $p$ labels the planet and $\Delta v_p(t)$ is the perturbation contributed by planet $p$. The perturbation is given by
\begin{equation}
\Delta v_p(t)=A_p[\sin{(f_p+\varpi_p)}+e_p\sin{\varpi_p}],
 \label{eq:radv}
\end{equation}
where $A$ is the velocity amplitude, $\varpi$ is the longitude of periastron, $e$ is eccentricity and $f$ is the true anomaly, which depends on time. 
The amplitude $A$ can be given by
\begin{equation}
A=\frac{m_p}{m_s+m_p}\frac{\omega a \sin{i}}{\sqrt{1-e^2}},
\end{equation}
where $m_p$ is the mass of the planet, $m_s$ is the mass of the star,
$\omega$ is the mean angular velocity ($2\pi/P$, where $P$ is the
period), $a$ is the distance between the star and the planet and $i$ is the inclination between
normal direction of the orbit and the line of sight. The true anomaly
$f$ is related to the eccentric anomaly $E$ through
\begin{equation}
\cos{f}=\frac{\cos{E}-e}{1-e\cos{E}}
 \label{eq:trueanomaly}
\end{equation}
and $E$ is related to the mean anomaly $M$ through
\begin{equation}
M=E-e\sin{E}
 \label{eq:meananomaly}
\end{equation}
where $M=\omega t + \phi$, $\phi$ is the phase of pericenter passage. Putting the equations above together, we get
\begin{equation}
\Delta v_p(t)=A_p[\sin{(\psi(\omega_p t+\phi_p,e_p)+\varpi_p)}+e_p\sin{\varpi_p}],
 \label{eq:radvel}
\end{equation}
where $\psi$ is a function determined by (\ref{eq:trueanomaly}) and 
(\ref{eq:meananomaly}).

There are 5 independent parameters for each planet in
(\ref{eq:radvel}), which are $A_p$, $\omega_p$, $\phi_p$, $e_p$
and $\varpi_p$.  We use an equivalent set of parameters, $\omega$,
$A_c = \sqrt{A}\cos{\phi}$, $A_s = \sqrt{A}\sin{\phi}$, $e_c = \sqrt{e}\cos{\varpi}$, 
and $e_s = \sqrt{e}\sin{\varpi}$.  Although our MCMC method
is not effected by linear changes of variables, this nonlinear change
of variables did improve its performance.  We believe this is because
it bounds $\phi$ and $\varpi$ between $0$ and $2\pi$ periodically,
thus making full use of machine precision. Another possible change of parameters
is replacing $\phi$ in $A_c$ and $A_s$ with $\phi+\varpi$, leaving
the others ($\omega$, $e_c$, $e_s$) unchanged. This re-parametrization makes the joint phase
uncertainty in $\phi$ and $\varpi$ less nonlinear, because $\phi$ and $\varpi$
are linearly correlated, see \figurename~\ref{fig:ep}.
However, any nonlinear change of variables in parameter space
introduces a Jacobian factor in the prior and likelihood. Replacing
$\phi$ in $A_c$ and $A_s$ with $\phi+\varpi$ causes a complicated Jacobian factor,
while for $\omega$, $A_c$, $A_s$, $e_c$ and $e_s$, which is the re-parametrization we
finally used, the Jacobian factor is just $1$.
For $n$ planets, there are $5n$ parameters.  Our model has two additional 
parameters.  One is the reference velocity offset $v_0$.  The other is the 
jitter, $s$, described in the following paragraph. 
(If there are multiple observatories or instruments, there will be multiple  
velocity offsets and jitters, which can be easily included in the code.)     
Together, these form the $M = 5n+2$ components of our parameter vector, 
$\vec{\theta}\in {\mathbb R}^M$.


The data for an estimation consists of $N$ observations of the radial
velocity of a given star.  For observation $i$, $t_i$ ($\unit{MJD}$ in
$\unit{d}$) is the observation time, $v_i$ ($\unit{m}\,\unit{s}^{-1}$) is the
observed radial velocity, and $\sigma_i$ ($\unit{m}\,\unit{s}^{-1}$)
is the reported standard deviation of the observation error.  This is
the observer's estimate of the measurement noise, induced by the instrument.
It does not include astrophysical noise
such as stellar oscillation.  We
denote the data collectively as $D = ( t_1,v_1,\sigma_1,\ldots,
t_N,v_N,\sigma_N)$.  Our data model assumes that observed velocities
are independent and Gaussian with mean $v_{\mbox{\scriptsize\em
    rad}}(t_i)$ and variance $\sigma_i^2 + s^2$.  The parameter
$s$ is the jitter referred to above which consists of all the noises not
included in the measurement noise estimation $\sigma$.  
Jitter increases the posterior
uncertainties of the parameters, so the parameter estimation is less optimistic
and thus more conservative \citep{gregory05a}.  The probability
density to observe data $D$, given parameters $\vec{\theta}$,
therefore is given by the Gaussian likelihood function
\begin{equation}
p(D|\vec{\theta})=
(2\pi)^{N/2}\bigg[\prod_{i=1}^N(\sigma_i^2+s^2)^{-1/2}\bigg]
\exp{\Bigg[-\sum_{i=1}^{N}\frac{(v_i-v_{rad}(t_i))^2}{2(\sigma_i^2+s^2)}\Bigg]} \; .
\label{pd}  \end{equation}
The parameters $\vec{\theta}$ enter into this formula through the dependence of 
$v_{\mbox{\scriptsize\em rad}}$ on the parameters.
It is helpful to write this in terms of the negative log of the denormalized
likelihood function
\begin{equation}
l(D|\vec{\theta}) = \sum_{i=1}^{N}\frac{1}{2} \left [
     \frac{(v_i-v_{rad}(t_i))^2}{2(\sigma_i^2+s^2)} +
     \log\left( \sigma_i^2+s^2) \right) \right] \; .
\label{ll} \end{equation}
And the likelihood function is $p(D|\vec{\theta})\propto e^{-l(D|\vec{\theta})}$.

We assume that the parameter vector $\vec{\theta}$ has a prior
distribution in which the individual parameters are independent random
variables whose distribution is given in \tablename~\ref{tab:prior}. For
demonstration purposes, we used a rather conventional non-informative
prior. Planet identifications would be more reliable with a prior that better
reflects our present understanding and expectations about the statistics of
exoplanets. We will leave that as a topic for future work.
The prior density, $p(\vec{\theta})$, is the product of the
appropriate densities over the $M = 5n+2$ components of
$\vec{\theta}$.  The posterior density of $\vec{\theta}$ given the
data is
\begin{equation}
p(\vec{\theta}|D)=\frac{p(\vec{\theta})p(D|\vec{\theta})}{p(D)} \; .
\end{equation}
In the present work, the normalization constant
$$
p(D) = \int p(\vec{\theta})p(D|\vec{\theta}) \,d\vec{\theta}
$$
is unknown and not used.

\section{Sampling}

\paragraph{Affine Invariance}
We use an ensemble sampling method that respects affine invariance. 
An affine transformation is an invertible mapping $\mathbb{R}^M\mapsto\mathbb{R}^M$ 
with the form $\vec{\psi}=S\vec{\theta}+b$ where $S$ is a non-singular matrix. 
Many Monte Carlo methods use diagonal or off diagonal scaling matrices $S$
to improve performance.
That is irrelevant for affine invariant methods.
If random variable $\vec{\theta}$ has probability density $\pi(\vec{\theta})$, 
then $\vec{\psi}=S\vec{\theta}+b$ has density
\begin{equation}
\pi_{S,b}(\vec{\psi})=\pi_{S,b}(S\vec{\theta}+b)\propto\pi(\vec{\theta}).
\end{equation}
A sampler is affine invariant if the MCMC transition probability density
$P(\vec{\theta}\to \vec{\theta}^{\prime})=P(\vec{\theta}^{\prime}\mid \vec{\theta})$ 
transforms in the same way:
$$
P(\vec{\psi}^{\prime}\mid \vec{\psi}) = 
C_{S,b}P(\vec{\theta}^{\prime}\mid \vec{\theta}) \; ,
$$
if $\vec{\psi}^{\prime}=S\vec{\theta}^{\prime}+b$ and $\vec{\psi}=S\vec{\theta}+b$,
where $C_{S,b}$ is a normalizing determinant independent of $\vec{\theta}$ and 
$\vec{\theta}^{\prime}$.
In our codes, this is a consequence of the fact that if the code starting with 
$\vec{\theta}_1$ would produce $\vec{\theta}_2, \vec{\theta}_3,\ldots$,
then starting with $\vec{\psi}_1$, it would produce $\vec{\psi}_2, \vec{\psi}_3,\ldots$,
always with $\vec{\psi}_j = S \vec{\theta}_j + b$.

Common MCMC samplers require certain customization to sample ill-shaped densities 
efficiently \citep{ford06a, gregory11a}.  For the density such as the left
side of \figurename~\ref{fig:ep}, single variable Gibbs sampler updates and
isotropic Metropolis samplers require a small step size to achieve a
reasonable acceptance probability.  But an affine transformation can
turn the highly elliptical density on the left to something more
spherical, as on the right.  The affine invariant MCMC sampler does not view
the ill-shaped density on the left more difficult than the well-shaped
density on the right\citep{goodman10a}.  This makes it
efficient on many ill-shaped densities, without customization.

\paragraph{Ensemble Sampling}
We use an ensemble sampler that updates $L$ samples together.  An
ensemble, $E$, consists of $L$ samples, or walkers, $\vec{\theta}_k$.
The walkers are samples of the posterior parameter distribution,
$\vec{\theta}_k \in {\mathbb R}^M$.  The ensemble $E =
(\vec{\theta}_1,\ldots,\vec{\theta}_L)$ of $L$ walkers is a point in
${\mathbb R}^{ML}$.  The target probability density for the ensemble
is
\begin{equation}
\Pi(E)=p(\vec{\theta_1}|D)p(\vec{\theta_2}|D)\cdots p(\vec{\theta_L}|D).
\label{PI}  \end{equation}
That is to say that each walker is an independent sampler of the posterior.

\paragraph{Stretch Move}
In the ensemble MCMC sampler, one step of the Markov chain $E(t)\to
E(t+1)$ consists of updating all walkers one by one as a whole
cycle. Expressed in pseudo code, it is
\begin{sffamily}
\begin{itemize}
\item[for] $k=1,\ldots,L$\\
update $\vec{\theta}_k(t)\to\vec{\theta}_k(t+1)$
\end{itemize}
\end{sffamily}
The update step preserves the ensemble sampling density (\ref{PI}), which means that 
if 
$$
E_{[k-1]}(t) = \left( \vec{\theta}_1(t+1),\ldots,\vec{\theta}_{k-1}(t+1), 
                    \vec{\theta}_k(t),\ldots,\vec{\theta}_L(t)           \right) \; ,
$$
has density (\ref{PI}), 
then $E_{[k]}(t)$ has density (\ref{PI}) too.
The update of walker $\vec{\theta}_k$ uses the current positions of the other walkers
in the ensemble, which we call the complementary ensemble:
$$
F_{[k]}(t) = \left\{ \vec{\theta}_1(t+1),\ldots,\vec{\theta}_{k-1}(t+1), 
                     \vec{\theta}_{k+1}(t),\ldots,\vec{\theta}_L(t)         \right\} \; .
$$
This is $E$ with its most current locations, but with walker $\vec{\theta}_k$ omitted.

In this work, we update $\vec{\theta}_k$ using the {\em stretch move}, defined as 
follows.
The random stretch variable, $Z$, has density $g(z)$ described below.
First, choose another walker $\vec{\theta}_j \in F_{[k]}(t)$, with all walkers 
in $F_{[k]}(t)$ being equally likely.
Then propose a move $\vec{\theta}_k \to \vec{\psi} = Z  (\vec{\theta}_k -\vec{\theta}_j)$.
The proposed new position of $\vec{\theta}_k$ lies on the line containing $\vec{\theta}_k$
and $\vec{\theta}_j$, but the distance between them is stretched by $Z$.
Accept $\vec{\psi}$ with Metropolis probability 
\begin{equation}
\mathrm{min}\Bigg\{1,Z^{M-1}\frac{p(\vec{\psi}|D)}{p(\vec{\theta}_k(t)|D)}\Bigg\}\; ,
 \label{fm:acceptanceprob}
\end{equation}
where $M$ is the dimension of the parameter space.
If $\vec{\psi}$ is accepted, then $\vec{\theta}_k(t+1) = \vec{\psi}$.
Otherwise  $\vec{\theta}_k(t+1) =\vec{\theta}_k(t)$.

The density of the stretch factor, $g(z)$, must satisfy the symmetry condition
\begin{equation}\label{eq:symmetrycondition}
g\left(\frac{1}{z}\right)=z\,g(z).
\end{equation}
As long as (\ref{eq:symmetrycondition}) is satisfied, the move 
$E_{[k-1]}(t) \to E_{[k]}(t)$ satisfies
detailed balance for the density (\ref{PI}), see \citep{goodman10a}.
A simple distribution that satisfies this condition is
\begin{equation}\label{fm:gz}
g(z)=\left\{
\begin{matrix}
\frac{1}{C}\,\frac{1}{\sqrt{z}}, & z\in\left[\frac{1}{a},a\right],\\
0, & otherwise.
\end{matrix}\right.
\end{equation}
The two parameters in the stretch move ensemble sampler are the ensemble size, $L$, 
and the stretch factor parameter, $a$.
The normalization constant is $C = \frac{1}{2}\left( \sqrt{a} - \frac{1}{\sqrt{a}}\right)$.
The method requires $L > M$ in order to sample the entire parameter space.
In the runs reported here, we took $L = 1000$ and $a = 2$.
The results are insensitive to ensemble size.
The parameters are not tuned for individual star datasets.


In pseudo code, moving $\vec{\theta}(t)\to\vec{\theta}(t+1)$ by one stretch move is given by \citep{goodman10a}
\begin{sffamily}
\begin{itemize}
\item[for] $k=1,\ldots,L$\\
choose $j \in \left\{1,\ldots,L\right\}$ randomly, with $j \neq k$\\
generate $\vec{\psi}=\vec{\theta}_j+Z(\vec{\theta}_k-\vec{\theta}_j)$, $Z$ satisfying (\ref{fm:gz})\\
accept, set $\vec{\theta}_k=\vec{\psi}$, with probability (\ref{fm:acceptanceprob})\\
otherwise reject, leave $\vec{\theta}_k$ unchanged.
\end{itemize}
\end{sffamily}

\section{Local Optima}
\begin{figure}
 \includegraphics[width=0.99\linewidth]{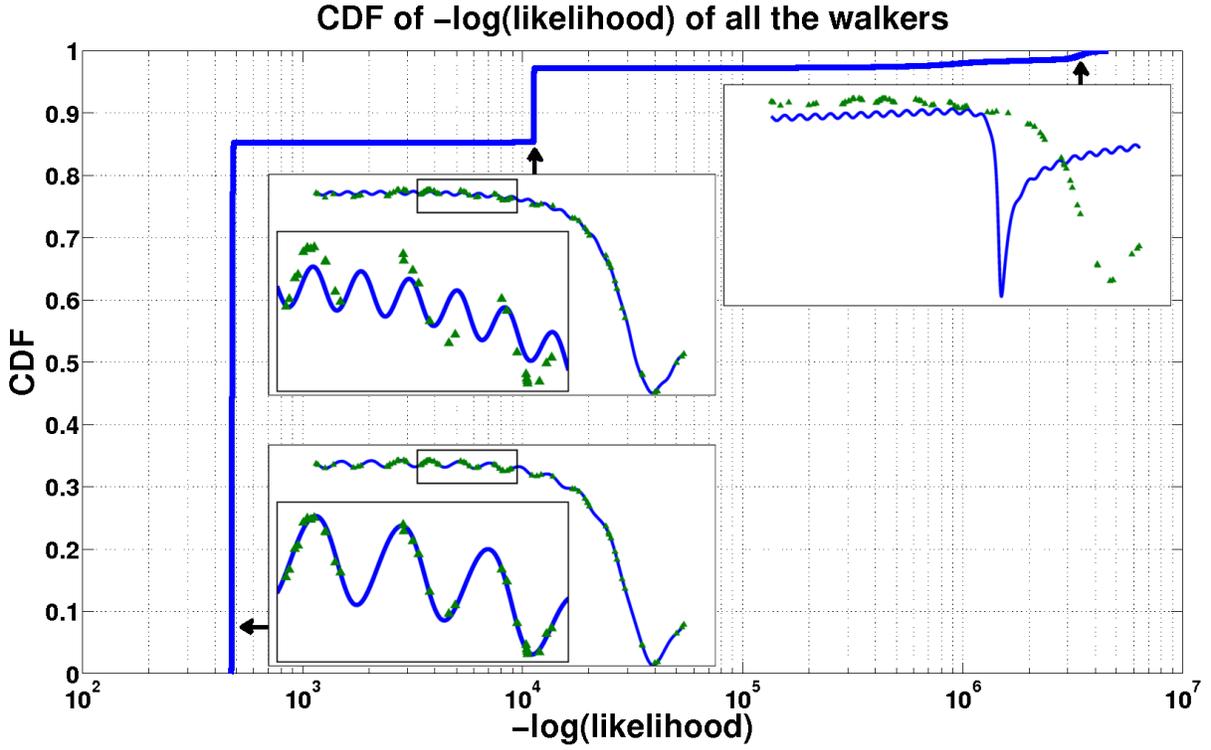}
 \caption{CDF of (\ref{lbar}), looking for $n=2$ companions in the RV
   data for (HIP36616).  There are $M=2 \cdot 5 + 2$ parameters.  The
   ensemble size was 1000 and the $\overline{l}_k$ were computed by
   averaging over $T=100$ samples.  The inserts illustrate the fits
   corresponding to the best-fit optimum and two local optima.}
\label{fig:loglikelihoodcdf}
\end{figure}

Local optima of the negative log likelihood function (\ref{ll}) are a
problem for many parameter estimation and sampling problems.  Local
optima correspond to parameter fits that are better than other nearby
fits but worse than fits in different parts of parameter space.
\figurename~\ref{fig:loglikelihoodcdf} gives an example for one specific
star dataset.  While any valid MCMC sampler would eventually find
samples near the global optimum, doing so may require an impractically
long run.  With the initialization just described, after cooling (see below), walkers   
from the ensemble find themselves in various local probability wells.
The computational challenge is twofold.  We seek to explore parameter
space well enough that at least some of the walkers find the well of
the global optimum.  
We also seek to
cluster walkers in the various wells so that we can remove walkers from 
all but the most important one, or ones. 


It is clearly undesirable to manually initialize the MCMC code star-by-star to
avoid local optima.
To be most useful, a code should find samples near the global optimum
by itself.  There seems to be no practical way to guarantee this, but
our combination of ensemble sampling with simulated annealing and
clustering does it more reliably than our earlier codes.  Also,
the faster equilibration time of the ensemble sampler allows it to
move more easily from local well to local well.  For HIP36616, we believe
our algorithm has identified the global optimum. However, in other datasets,
our algorithm failed to find fits as good as those other groups found by hand
using the periodogram.  Further progress on this issue would be very important
in creating automatic data analysis routines that do not require individual
human attention for each star.  For instance, right now, we only use the periodogram
to roughly guess the period (see below), but we could make further use of the periodogram
in our code to improve automatization.

\paragraph{Simulated Annealing}
Simulated annealing \citep{simul} is a Monte Carlo method for finding global optima
in problems that have many local optima.
It uses a modified posterior distribution with an artificial ``inverse temperature''
parameter, $\beta$:
\begin{equation}
p_{\beta}(\vec{\theta}|D) \propto p(\vec{\theta})e^{-\beta l(D|\vec{\theta})} \; ,
\label{beta}  \end{equation}
where $l$ is the negative log likelihood function (\ref{ll}).  Small
$\beta$ allows the sampler to explore parameter space freely without
getting stuck in local wells.  The desired posterior (\ref{pd})
corresponds to $\beta = 1$.  The autocovariance function, described
below, decays faster at small $\beta$.  Cooling (increasing $\beta$)
slowly allows the walkers to find promising local optima.  A cooling
schedule is an algorithm that increases $\beta$ during the sampling
process.

\begin{figure}
 \includegraphics[width=0.9\linewidth]{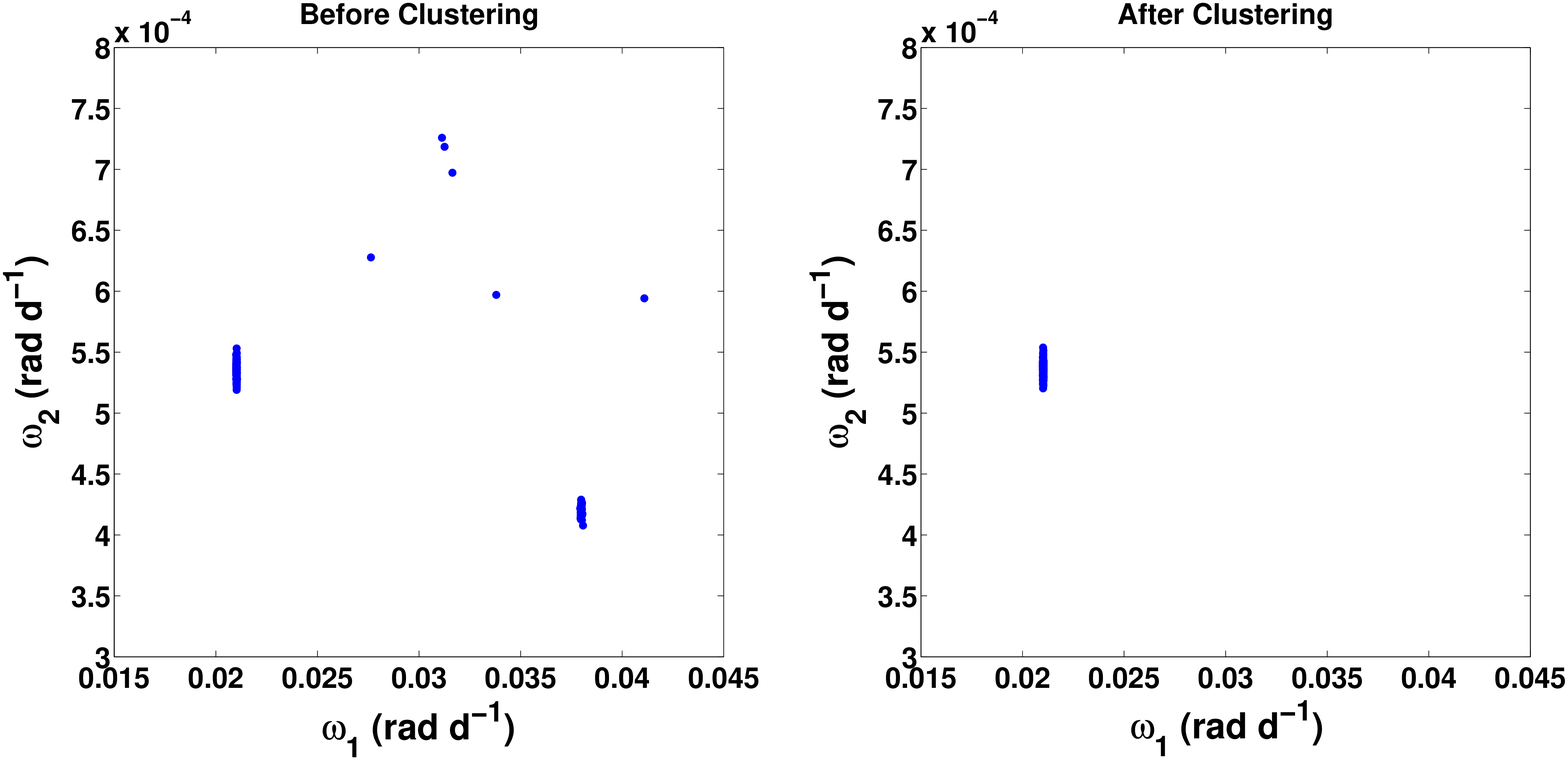}
 \caption{Two dimensional scatterplots of the walkers in the size
   $L=1000$ ensemble of \figurename~\ref{fig:loglikelihoodcdf}.  Plotted
   are the angular velocity parameters for the two companions.  The left
   frame contains all the walkers in the ensemble after annealing.
   The right frame contains only those walkers that the code
   identified as being in the deepest well, corresponding to the best
   fit.}
\label{fig:clustering}
\end{figure}

We use simulated annealing in the first phase of our sampler.  We initialize 
the ensemble as a small ball centered at a random position in parameter space 
except for the dimensions related to the period, for which we use the periodogram 
\citep{bretthorst88a, cumming04a} to check for peaks between the shortest
observation interval and the total observation time to make a guess. As the dataset can reveal
periods outside this range, this is not completely reliable. In a one-planet fit, we use
the period with the highest peak $P_{highest}$ as the initial period guess unless 
the data shows a linear trend which indicates a long-period companion. In such 
cases we use $P_{highest}/100$ as the period guess. In two-planet fit, 
$P_{highest}$ is the initial guess for the 1st planet and for the 2nd planet, 
we use the period of the 2nd highest peak to initialize a portion of walkers  
and the period of the 3rd highest peak for another portion of walkers and so on. 
If the data shows a linear trend, we would use $P_{highest}/100$
as the guess for the 2nd planet. The number $100$ is chosen empirically and any
reasonably large number should do. Ideally, we want our initialization to be as
random as possible to demonstrate the value of the algorithm, but not as random so that the
algorithm is unnecessarily sabotaged by the randomness.
We initialize $\beta$ by $\beta_{min} = 1/N$, where $N$ is the size of dataset 
so that only one data matters on average in the beginning. 
The runs here used the following cooling schedule: $\beta$ is increased in $N-1$ steps
and the amount is determined by $1/\beta_i - 1/\beta_{i+1} = 1/(N\beta_{min})$
where $i$ is the index of the steps. $\beta$ is kept constant within every 
one $N$th of the total steps. In general,  we used $N = 10$.  At the end of 
the cooling phase the sampler has been using $\beta = 1$ for a while,
so the walkers in the ensemble are well equilibrated in their local optima.
A typical example is the left frame of \figurename~\ref{fig:clustering},
with several local optima clearly visible.






\paragraph{Clustering}
\figurename~\ref{fig:loglikelihoodcdf} shows the results for a
particular dataset after annealing, for which about $85\%$ of the
walkers in the ensemble are in a well
that corresponds to an excellent fit.  About $12\%$ of the remaining walkers
are in another local well that corresponds to a worse fit (see inserts).  The negative log likelihood
for this worse fit is an order of magnitude larger.  The clustering phase of our
algorithm seeks to identify clusters of walkers corresponding to the
local wells, so that all but the important ones can be removed.  The
right frame of \figurename~\ref{fig:loglikelihoodcdf} shows the result
in this case.  Only walkers from the global optimum remain. 

While there are many sophisticated clustering methods, we found that for our 
problems, a simple one dimensional clustering method is more effective and 
reliable than the others we tried, such as clustering in parameter space. 
Note that there could also be multiple local optima due to aliasing, they could 
have very similar likelihoods. In such cases, we would need more sophisticated  
clustering methods.                                                             
After the annealing is finished, but before clustering, we collect the average 
negative log likelihood function for each walker.
(One can also use posterior function, since the normalization constant doesn't 
matter when one is taking the difference of posterior.)                        

This results in the $L$ numbers
\begin{equation}
\overline{l}_k = \frac{1}{T} \sum_{t=1}^T l(\vec{\theta}_k(t)|D) \; .
\label{lbar}  
\end{equation}
The idea is that $\overline{l}_k$ is characteristic of the well walker
$\vec{\theta}_k$ is in, so that walkers in the same well will have
similar $\overline{l}_k$.  Given $\overline{l}_k$, the clustering
first ranks all the walkers based on $\overline{l}_k$ so that
$\{\vec{\theta}_{(1)},\vec{\theta}_{(2)},\ldots,\vec{\theta}_{(L)}\}$
is in the order of decreasing $\overline{l}_{(k)}$, or increasing
$-\log{\overline{l}_{(k)}}$.  There are big jumps in
$-\log{\overline{l}_{(k)}}$ in this sequence (see
\figurename~\ref{fig:loglikelihoodcdf}), which are fairly easy to
identify and thus to separate the jumps. We calculated the difference in
$-\log{\overline{l}_{(k)}}$ for every adjacent pair of
$\vec{\theta}_{(k)}$ starting from $k=1$ and find the 1st pair whose
difference is certain amount of times bigger than the average
difference before. So if
\begin{equation}
- \log{\overline{l}_{(j+1)}} + \log{\overline{l}_{(j)}} > Const \frac{- \log{\overline{l}_{(j)}} + \log{\overline{l}_{(1)}}}{j-1},
\end{equation} 
all the $\vec{\theta}_{(k)}$ with $k > j$ are thrown away and only the ones 
with $k \leq j$ are kept.

A more rigorous approach to the multiple well/multiple fit problem would be to 
estimate the evidence integral corresponding to a well before deleting it.  It 
could happen that a small but deep well has less posterior probability than a 
broad but shallow well.  We do not find that this occurs in practice in the fits discussed in 
this paper.  For example, our prior prevents very short-period orbits that might give 
excellent but spurious fits to the data.  The inserts of 
\figurename~\ref{fig:loglikelihoodcdf} suggest that the local wells should be 
removed in our problem.  In future work we hope to address this issue more 
carefully by estimating the evidence integrals.  Even then, clustering as we do 
it here will be the first step.



\section{Data and results}

We tested our code on data from the Lick K-Giant Search \citep{frink02a, mitchell03a, hekker06a, hekker08a, quirrenbach11a}. 
In this paper, we present the results for two stars: HIP36616 and HIP88048. 
HIP36616 is interesting because the data shows
a companion with a small period and small mass and
another companion with very large period and very large mass. The information
on the large period is incomplete which causes the Metropolis-Hastings algorithm difficulty in finding
a good fit. HIP88048 has near complete information on both periods so it is an
easier case for Metropolis Hastings and is used as a comparison. Other groups ran these two stars
with common Metropolis-Hastings routines. They confirmed that they could easily find a good fit
for HIP88048. But for HIP36616, a good fit was found only after new data were obtained. Our code
successfully found a fit for HIP36616 before the new data came in.

\begin{figure}
 \includegraphics[width=0.96\linewidth]{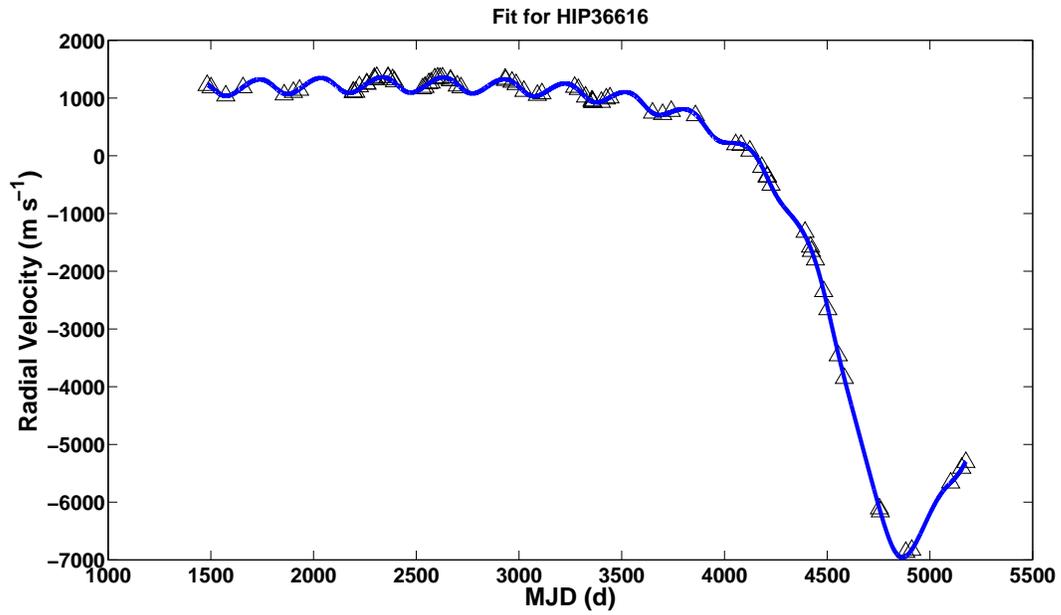}
 \caption{The fit of HIP36616. See also \figurename~
   \ref{fig:loglikelihoodcdf} for worse fits that sit in deep local
   wells.}
\label{fig:fit122}
\end{figure}

\begin{figure}
 \includegraphics[width=0.9\linewidth]{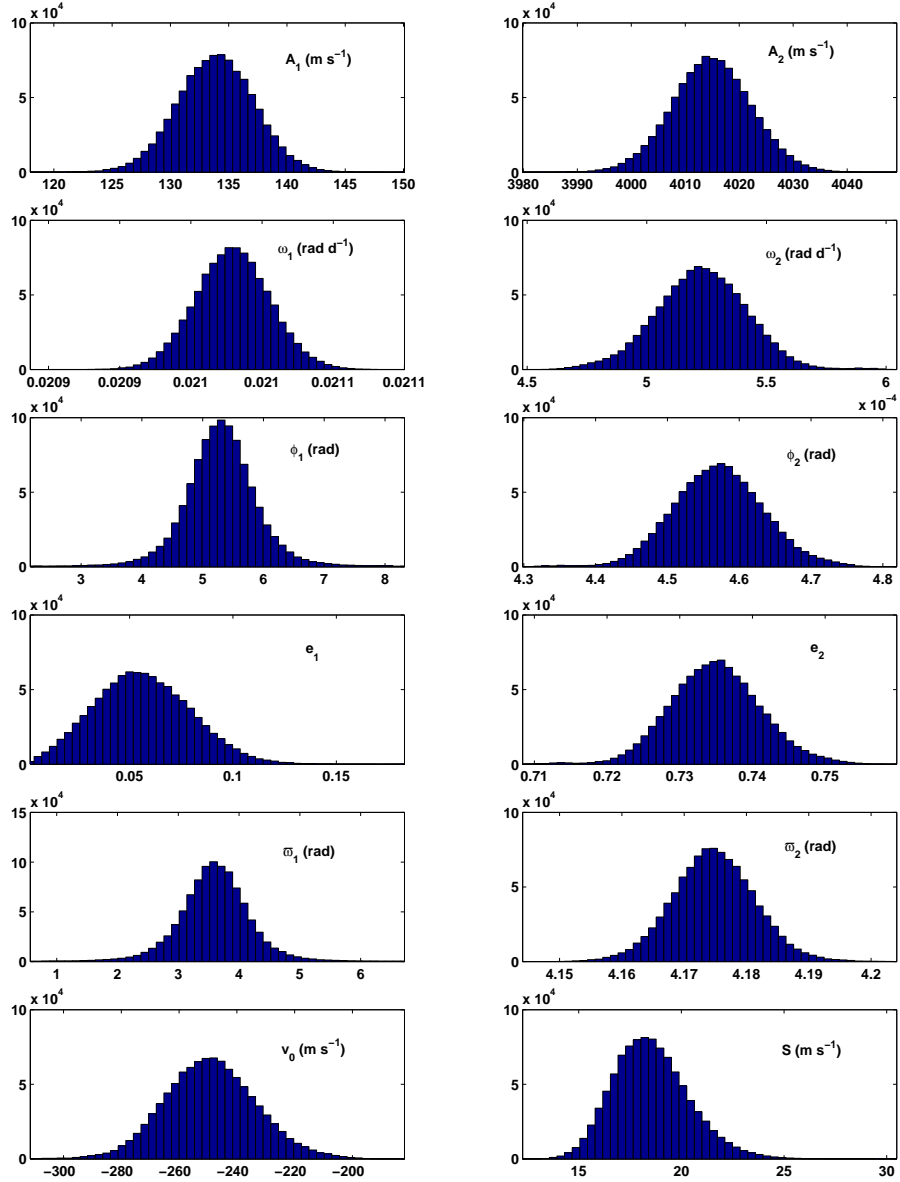}
 \caption{The histograms of the parameter posteriors from HIP36616.}
\label{fig:pdf122}
\end{figure}

HIP36616 is interesting and challenging both from a physical and a
sampling point of view.  The RV data together with a good fit are
shown in \figurename~\ref{fig:fit122}.  The data are well explained by a
small close companion in a roughly circular orbit and a larger and more
distant companion in a highly elliptical orbit.  The mass ratio in the
posterior distribution (assuming coplanar orbits) is
$$
\frac{m_{p2}}{m_{p1}}=\frac{A_2}{A_1}\frac{\sqrt{1-e_2^2}}{\sqrt{1-e_1^2}}\frac{\omega_1^{1/3}}{\omega_2^{1/3}}
=69.8\pm1.7 .
$$
The parameters with their confidence intervals are listed in \tablename~\ref{tab:122param}.
Histograms of the individual parameters are shown in \figurename~\ref{fig:pdf122}.

\begin{figure}
 \includegraphics[width=0.96\linewidth]{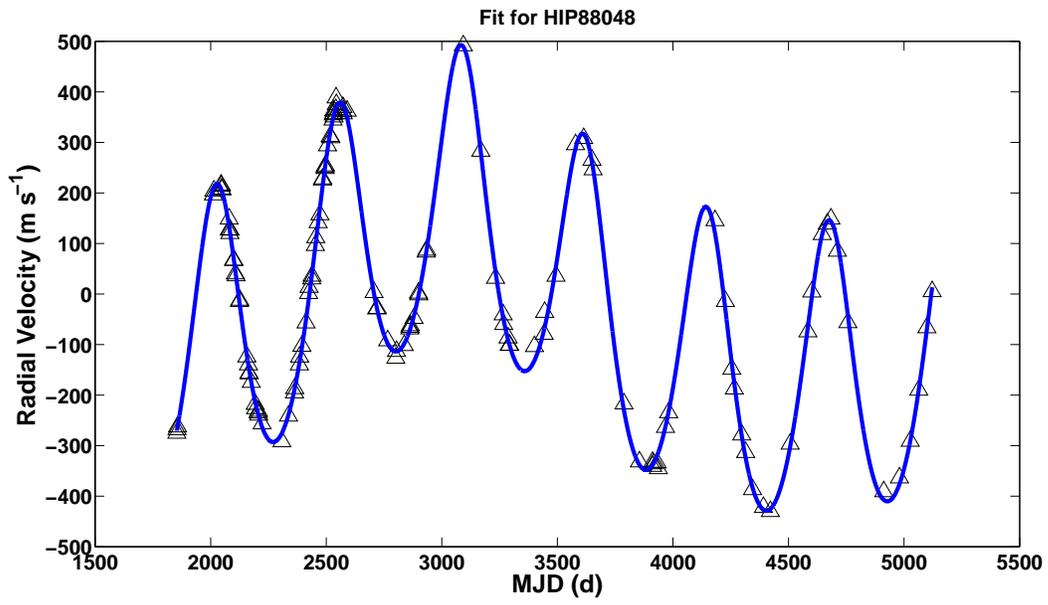}
 \caption{The fit of HIP88048}
\label{fig:fit282}
\end{figure}

The second example fit with the new sampler, HIP88048, is shown in  \figurename~\ref{fig:fit282}.
The parameters with their confidence intervals are listed in \tablename~\ref{tab:282param}.
We also run our code on only the first half of the data of HIP88048. We get similar   
results as with all the data, but with a larger variance, especially for the   
parameters of the planet with larger period.                                   

\section{Comparison with Metropolis-Hastings}

\begin{figure}
\includegraphics[width=0.99\linewidth]{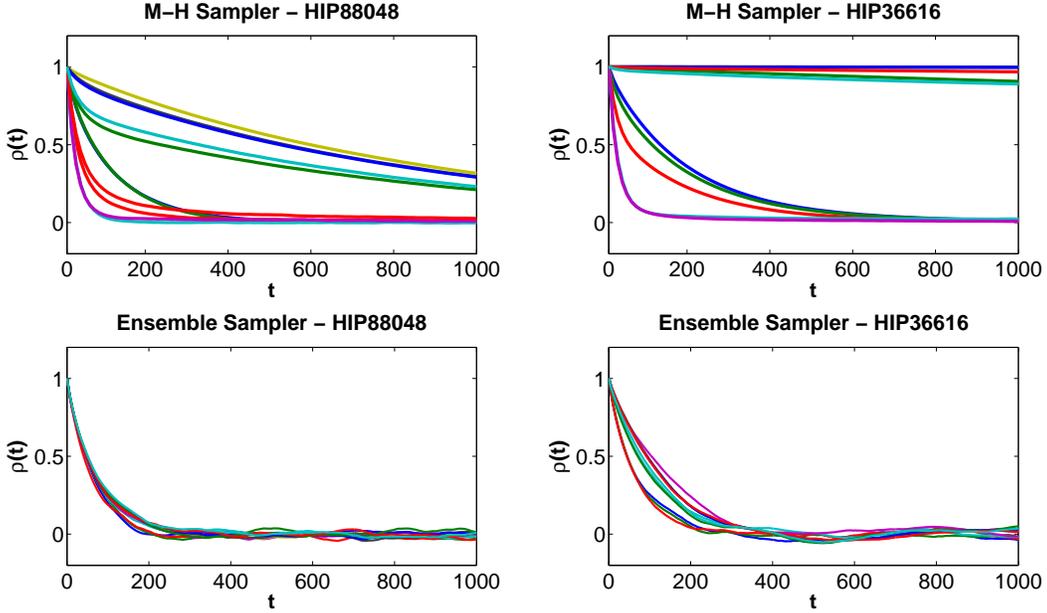}
\caption{After convergence is achieved, we take chains of all the parameters to
  calculate the autocorrelation function which shows how many steps it takes
  to generate an independent sample. The autocorrelation functions for the
  traditional Metropolis-Hastings sampler are the 1st row and the 2nd row is for
  the Affine Invariant Ensemble Sampler.  Data are taken after the cooling and 
  clustering are completed.  All of the runs used $10^7$ resamplings.  The left
  two frames are for HIP88048, which is relatively unambiguous and
  easier to fit.  The two right frames are for HIP36616, which is harder. Both
  sampler uesd the same parametrization described in the paper. For Metropolis-
  Hastings sampler, we tuned all the parameters so that the acceptance rate is 
  around 0.4 in all the dimensions. We did not explore all the non-linear 
  transformations of the parameters nor all the proposed PDFs, but this is what 
  most people do. For our ensemble sampler, we also tuned the acceptance rate to
  be around 0.4. }
\label{fig:ac}
\end{figure}

All non-trivial MCMC samplers produce autocorrelated samples.  An important measure of
the effectiveness of the sampler is its covariance $C(t)$ of the 
equilibrium time lag $t$.  The longer it takes the covariance $C(t)$ to decay to $0$,
the longer it takes the sampler to generate independent samples, 
because non-zero covariance indicates correlation between samples. 
To be more precise, suppose $V(\vec{\theta})$ is
some function of the parameters, such as $V(\vec{\theta}) = \theta_j$
or some nonlinear function of the components.  The equilibrium
autocovariance function is
$$
C_V(t) = \lim_{t_0\to\infty} 
   \mbox{cov}\!\left[  
      V(\vec{\theta}(t_0+t)), V(\vec{\theta}(t_0))\right] \; .
$$
The limit $t_0 \to \infty$ is only to ensure that the Monte Carlo has 
reached steady state.
The dimensionless version of this is the autocorrelation function 
\begin{equation} 
\rho_V(t) = \frac{C_V(t)}{C_V(0)} = 
\lim_{t_0 \to \infty} 
\frac{\mbox{cov}\!\left[V(\vec{\theta}(t_0+t)), V(\vec{\theta}(t_0))\right] }
     {\mbox{var}\!\left[V(\vec{\theta}(t_0) \right]} \; .
\label{rho}
\end{equation}
We used the standard estimators
$$
\widehat{C}_V(t) \;=\;
\frac{1}{N-t}\sum_{n=0}^{N-t}{(V(\vec{\theta}(n+t))-\overline{V})
                              (V(\vec{\theta}(n))-\overline{V})} \;,
$$
and
$$
\widehat{\rho}_V(t) \;=\; \frac{\widehat{C}_V(t)}{\widehat{C}_V(0)} \; .
$$
\figurename~\ref{fig:ac}  displays the functions $\widehat{\rho}_j(t)$, which correspond to 
taking $V(\vec{\theta})$ to be the j-th parameter $\theta_j$.
Note $\rho_j(t)$ can be quite different for different
components of $\vec{\theta}$ for the same dataset.

The important measure of correlation
for Monte Carlo is the integrated autocorrelation time (summed,
actually)
$$
\tau = \sum_{t=-\infty}^{\infty} \rho(t)  = 1 + 2\sum_{t=1}^{\infty} \rho(t) \; .
$$
This is the number of Monte Carlo steps needed to produce an effectively independent
sample of the posterior, see \citep{goodman10a} and references there.

\paragraph{Metropolis Sampler Proposal and Tuning.} 

For comparison purposes we coded a traditional MCMC sampler that
updates the components $\theta_j$ one at a time using a trial that is
uniform in an interval $[\theta_j - r_j,\theta_j+r_j]$.  The ranges
$r_j$ were tuned individually for each component to give acceptance
probability $.4$, which is commonly thought to be roughly optimal.
The tuning parameters are different for each $j$, and for a given $j$
they are different for each dataset.
One can also use Gaussian distribution for trial move which should give similar  
result.                                                                          


It is noteworthy that for the Metropolis sampler to work best, one should tune
all the free parameters which are on the order of the dimension squared. But
how the free parameters are tuned highly depends on where the walker is
in the parameter space. That is to say, the free parameters best tuned for
a place far from the best-fit optimum in the parameter space could be quite
different from those best tuned for a place close to the best-fit optimum.
Even if the best-tuned free parameters are similar everywhere in the parameter
space, the tuning workload still increases as fast as the parameter space dimension
squared. 

\paragraph{Comparison.} 

For the ensemble sampler, we used the ensemble mean
\begin{equation}
\bar{V}(\vec{\theta}_n(t)) = \frac{1}{L} \sum_{n=1}^L V(\vec{\theta}_n(t))
\label{eq:ens_mean}
\end{equation}
in place of a single value on the right side of (\ref{rho}) in the
definition of $\rho$,
in other word, replacing $V$ in (\ref{rho}) with $\bar{V}$ defined in (\ref{eq:ens_mean})        
(Note the possible confusion of notation:
$\theta_n$ is component $n$ of the parameter vector $\vec{\theta}$,
while $\vec{\theta}_n$ is the $n-$th parameter vector in an ensemble
of $L$ such parameter vectors.)  It may seem unfair to compare one
step of a single vector method to a single step of the ensemble
method, given that an ensemble step requires updating $L$ walkers
rather than one.  However, because the walkers in the ensemble are
independent (see (~\ref{PI})), doing $\tau$ ensemble updates produces
$L$ effectively independent samples. \footnote{This does not apply to burn
in, so that ensemble may suffer more work from burn in phase.} Therefore,
the ensemble sampler will be exactly as effective
as the traditional one if it has the same autocorrelation function \citep{goodman10a} .



Two other factors should be mentioned.  One is that the traditional
sampler requires one likelihood evaluation per component per update,
while the ensemble sampler requires one likelihood evaluation per
vector update.  That is, the traditional sampler uses a factor of $M$
(the number of parameters) more likelihood evaluations per vector
update.  This is a serious consideration in the present application,
where likelihood evaluations are the most expensive part of the
algorithm.  To be sure, in many applications (though not ours) it is
not so expensive to update the likelihood function after changing a
single parameter.  The other is that the ensemble method is more
automatic in that it does not require component specific or dataset
specific tuning.  All the runs reported here used parameters $a=2$ and
$L=1000$.

\figurename~\ref{fig:ac} shows the results for the two datasets, each of which is a
star with two companions.  In both cases, the traditional sampler relaxes
the components of the easier companion more quickly, while the ensemble
sampler (which updates all parameters together) relaxes all parameters
at roughly the same rate.  In both cases, the ensemble is much more
effective in relaxing the parameters of the difficult companion.  In the
harder case, HIP36616, the relaxation for the ensemble sampler is at
least an order of magnitude faster than the relaxation for the
traditional sampler.

\section{Discussion}
This paper reports progress toward the goal of making posterior
sampling fast, reliable, and automatic for exoplanet radial velocity fitting. 
The sampler presented here performed well and automatically for
all the individual star datasets we tried (about a hundred data points). 
Very high aspect ratio wells and spurious local wells all were handled 
without individual tuning using a single shell script.
Future projects, such as hierarchical modeling of the exoplanet distribution \citep{hogg10a},
can become impractical with traditional samplers if they are slow.
Statistical information about exoplanets, such as the eccentricity
distribution \citep{shen08a, hogg10a, zakamska10a}, the mass distribution
\citep{howard10a}, the mass-semimajor axis distribution \citep{schlaufman09a}, 
the brown dwarf desert \citep{grether06a, leconte10a}, and inclinations \citep{ho10a},
depend on rapid and reliable processing of star datasets. 
It is a limitation of this work that we have used uninformative priors,
since we already have plenty of knowledge about exoplanets; we should infer from
the collection of all exoplanets priors that improve our fitting for any 
individual new system. Hierarchical approaches make it possible to infer 
informative priors without making strong new assumptions. Hierarchical model
comparison and selection is a future project, as are projects in which we tune
our optimization strategy to operate rapidly and robustly on large numbers of
exoplanet systems (a pre-requisite for efficient hierarchical modeling.

\begin{figure}
 \includegraphics[width=0.99\linewidth]{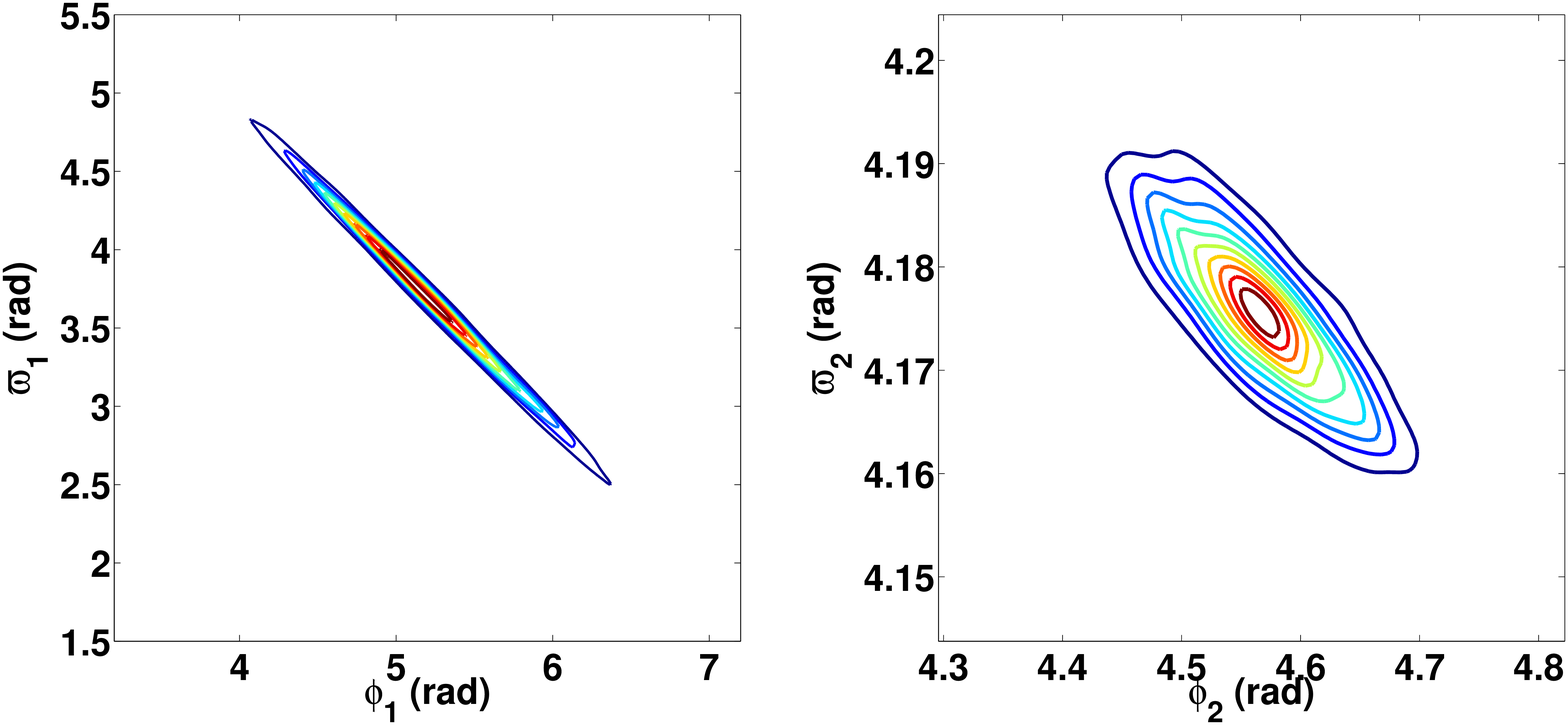}
 \caption{The relationship between $\phi$ and $\varpi$ of both
   companions of HIP36616.  We used Gaussian kernel density estimation
   to estimate the joint probability distribution of $\phi_1$ and
   $\varpi_1$ (left) and $\phi_2$ and $\varpi_2$ (right) from a large
   number of samples.}
 \label{fig:ep}
\end{figure}

We believe that \figurename~\ref{fig:ep} at least partly explains the
faster decay of correlations in the ensemble sampler.  For the
probability distribution in the left frame, traditional single
variable updates must have small proposal steps or suffer high
rejection rates.  For example, if $\phi_1$ is fixed, then $\varpi_1$
cannot move much and stay within the range of likely parameters.
Samplers based on isotropic multivariate proposals must also take
small proposal steps.  But the ensemble sampler moves walkers along
lines between themselves and other walkers, which naturally adapts the
proposal steps to the geometry of the distribution.

In our comparison with the Metropolis-Hastings sampler, we didn't tune the M-H sampler
to its full extent. If we did, it is reasonable to think that the performance
of the Metropolis-Hastings sampler---locally where it is tuned---would be the same
as our ensemble sampler. But tuning free parameters (whose number is on the order of
the number of dimensions squared) is already difficult, let alone the fact that best-tuned
free parameters could change dramatically as the walker travels in the
parameter space. Our ensemble sampler emulates a best-tuned Metropolis-Hastings sampler,
best-tuned wherever the walker goes. 

One of the problems we have encountered is that our ensemble sampler is
greatly affected by the existence multiple local optima. If the walkers occupy different
local optima, the acceptance ratio can be greatly lowered because
the proposed moves are based on walkers from other optima. 
In this case, the proposal may be outside any well and therefore unlikely to 
be accepted.
The clustering we propose and use here is a great help in this regard, and effectively overcomes this problem. And our runs never encountered more than one statistically
significant local optima.  


The ensemble sampler may also be better suited for high performance
computing.  Each walker in the ensemble evolves in parallel.  Our 
current implementations do not exploit this parallelism; more
sophisticated software designed for high performance hardware 
would be welcome. 

We believe that more sophisticated methods from machine learning could
have a large impact on these sampling problems.  One obvious
application would be more sophisticated clustering methods.  Another
would be to make more use of the information in the ensemble to build
a model of the posterior.  For example \citep{goodman10a} speculate
that one could use half of the ensemble to build a nonlinear model of
the posterior that would allow the other half of
the ensemble to be re-sampled more effectively

All of the software used for this project is available upon request.
Included in this is a general purpose of the ensemble sampler that can be used for
other sampling problems.

\acknowledgements It is a pleasure to thank Jo Bovy, Debra Fischer,
Daniel Foreman-Mackey and Ros Skeltonfor for useful discussions and
feedback.  

We want to particularly thank Andreas Quirrenbach for generously sharing data with us.

FH and DWH were partially supported by the NSF (grant
AST-0908357), NASA (grant NNX08AJ48G), and the Alexander von Humboldt
Foundation. JG and JW were partially supported by DOE grant DE-FG02-88ER25053.

\clearpage
\begin{table}
\centering
\begin{tabular}{cccc}
\hline
Parameter & Variable & Prior & Mathematical Form \\
\hline \noalign{\smallskip}
Amplitude & $A\,(\unit{m}\,\unit{s}^{-1})$ & Jefferys & $\frac{(A+A_0)^{-1}}{\ln{\big(\frac{A_0+A_{max}}{A_0}\big)}}$ \\   
        \noalign{\smallskip} \noalign{\smallskip}
Period & $P\,(\unit{s})=2\pi/\omega$ & Jefferys & $\frac{P^{-1}}{\log{(P_{max}/P_{min})}}$ \\  
           \noalign{\smallskip} \noalign{\smallskip}
Orbital Phase & $\phi\,(\unit{rad})$ & Uniform & $0 \leq \phi \leq 2\pi$ \\  
      \noalign{\smallskip} \noalign{\smallskip}
Eccentricity & $e$ & Uniform & $0 \leq e \leq 1$ \\ 
      \noalign{\smallskip} \noalign{\smallskip}
Longitude of Periastron & $\varpi\,(\unit{rad})$ & Uniform & $0 \leq \varpi \leq 2\pi$ \\
\noalign{\smallskip} \noalign{\smallskip}
Jitter & $s\,(\unit{m}\,\unit{s}^{-1})$ & Jefferys & $\frac{(s+s_0)^{-1}}{\ln{\big(\frac{s_0+s_{max}}{s_0}\big)}}$ \\ \noalign{\smallskip} \noalign{\smallskip}
Reference Velocity & $v_0\,(\unit{m}\,\unit{s}^{-1})$ & Uniform & 
 $v_{0\mbox{\scriptsize \em min}} \leq v_0 \leq v_{0\mbox{\scriptsize \em max}}$ \\ 
     \noalign{\smallskip}\hline \noalign{\smallskip}
\end{tabular}
\begin{center}
\caption{Priors for the model parameters.  The hyperparameters used in code are
$A_0 = 10\,\unit{m}\,\unit{s}^{-1}$, $s_0 = 10\,\unit{m}\,\unit{s}^{-1}$, $v_{0\mbox{\scriptsize \em min}} = -10000\,\unit{m}\,\unit{s}^{-1}$ and $v_{0\mbox{\scriptsize \em max}} = 10000\,\unit{m}\,\unit{s}^{-1}$.
$A_{max}$, $P_{max}$ and $s_{\mbox{\scriptsize\em max}}$ are in the normalization, so they are not actually used.\label{tab:prior}}
\end{center}
\end{table}

\clearpage
\begin{table}
\centering
\begin{tabular}{c|c|c|c}
\hline
parameter & median & $68\%$ CI & $95\%$ CI\\
\hline
$A_1(\unit{m}\,\unit{s}^{-1})$ & 133.9640 & $\pm 3.1450$ & $\pm 6.3495$\\
$\omega_1(\unit{rad}\,\unit{d}^{-1})$ & $2.10296\times 10^{-2}$ & $\pm 0.00259\times 10^{-2}$ & $\pm 0.00507\times 10^{-2}$\\
$\phi_1(\unit{rad})$ & 5.26268 & $\pm 0.52774$ & $\pm 1.24894$\\
$e_1$ & 0.0557283 & $\pm 0.0234535$ & $\pm 0.0452732$\\
$\varpi_1(\unit{rad})$ & 3.57290 & $\pm 0.51762$ & $\pm 1.24018$\\
$A_2(\unit{m}\,\unit{s}^{-1})$ & 4015.00 & $\pm 7.16$ & $\pm 14.22$\\
$\omega_2(\unit{rad}\,\unit{d}^{-1})$ & $5.23189\times 10^{-4}$ & $\pm 0.17656\times 10^{-4}$ & $\pm 0.35001\times 10^{-4}$\\
$\phi_2(\unit{rad})$ & 4.57002 & $\pm 0.05873$ & $\pm 0.11665$\\
$e_2$ & 0.734762 & $\pm 0.005779$ & $\pm 0.011469$\\
$\varpi_2(\unit{rad})$ & 4.17465 & $\pm 0.00629$ & $\pm 0.01270$\\
$S\,(\unit{m}\,\unit{s}^{-1})$ & 18.4083 & $\pm 1.7982$ & $\pm 3.6052$\\
$v_0(\unit{m}\,\unit{s}^{-1})$ & -249.559 & $\pm 14.560$ & $\pm 28.969$\\
\hline
\end{tabular}
\begin{center}
\caption{HIP36616, posterior means and confidence intervals for the 12 parameters.
Note one nearly circular orbit ($e_1$) and one very eccentric one ($e_2$).\label{tab:122param}}
\end{center}
\end{table}

\clearpage
\begin{table}
\centering
\begin{tabular}{c|c|c|c}
\hline
parameter & median & $68\%$ CI & $95\%$ CI\\
\hline
$A_1(\unit{m}\,\unit{s}^{-1})$ & 288.108 & $\pm 1.261$ & $\pm 2.464$\\
$\omega_1(\unit{rad}\,\unit{d}^{-1})$ & $1.18567\times 10^{-2}$ & $\pm 0.00048\times 10^{-2}$ & $\pm 0.00096\times 10^{-2}$\\
$\phi_1(\unit{rad})$ & 4.12983 & $\pm 0.03176$ & $\pm 0.06322$\\
$e_1$ & 0.129846 & $\pm 0.004529$ & $\pm 0.008910$\\
$\varpi_1(\unit{rad})$ & 1.73223 & $\pm 0.03175$ & $\pm 0.06283$\\
$A_2(\unit{m}\,\unit{s}^{-1})$ & 175.842 & $\pm 1.588$ & $\pm 3.143$\\
$\omega_2(\unit{rad}\,\unit{d}^{-1})$ & $1.95700\times 10^{-3}$ & $\pm 0.02129\times 10^{-3}$ & $\pm 0.04227\times 10^{-3}$\\
$\phi_2(\unit{rad})$ & 3.85943 & $\pm 0.04639$ & $\pm 0.09299$\\
$e_2$ & 0.194608 & $\pm 0.012026$ & $\pm 0.024027$\\
$\varpi_2(\unit{rad})$ & 1.76762 & $\pm 0.03943$ & $\pm 0.07846$\\
$S\,(\unit{m}\,\unit{s}^{-1})$ & 7.7662 & $\pm 0.7582$ & $\pm 1.5029$\\
$v_0(\unit{m}\,\unit{s}^{-1})$ & -39.1676 & $\pm 1.4873$ & $\pm 2.9545$\\
\hline
\end{tabular}
\begin{center}
\caption{Parameters for HIP88048\label{tab:282param}}
\end{center}
\end{table}

\clearpage

\clearpage

\clearpage

\clearpage

\clearpage

\clearpage



\clearpage

\end{document}